# Do CCSD and approximate CCSD-F12 variants converge to the same basis set limits? The case of atomization energies

Manoj K. Kesharwani,[1,2] Nitai Sylvetsky,[1] Andreas Köhn,[2] David P. Tew,[3] and Jan M. L. Martin[1,a)]

[1]Department of Organic Chemistry, Weizmann Institute of Science, 76100 Reḥovot, Israel

[2]Institute for Theoretical Chemistry, University of Stuttgart, 70569 Stuttgart, Germany

[3]Max Planck Institute for Solid State Research, Heisenbergstraße 1, 70569 Stuttgart, Germany
a)Corresponding author: gershom@weizmann.ac.il

**Abstract.** While the title question is a clear 'yes' from purely theoretical arguments, the case is less clear for practical calculations with finite (one-particle) basis sets. To shed further light on this issue, the convergence to the basis set limit of CCSD (coupled cluster theory with all single and double excitations) and of different approximate implementations of CCSD-F12 (explicitly correlated CCSD) have been investigated in detail for the W4-17 thermochemical benchmark. Near the CBS ([1-particle] complete basis set) limit, CCSD and CCSD(F12*) agree to within their respective uncertainties (about ±0.04 kcal/mol) due to residual basis set incompleteness error, but a nontrivial difference remains between CCSD-F12b and CCSD(F12*), which is roughly proportional to the degree of static correlation. The observed basis set convergence behavior results from the superposition of a rapidly converging, attractive, CCSD[F12]–CCSD-F12b difference (consisting mostly of third-order terms), and a more slowly converging, repulsive, fourth-order difference between CCSD(F12*) and CCSD[F12]. For accurate thermochemistry, we recommend CCSD(F12*) over CCSD-F12b if at all possible. There are some indications that the nZaPa family of basis sets exhibits somewhat smoother convergence than the correlation consistent family.

## INTRODUCTION

Explicitly correlated (R12 and F12) electron correlation methods (see, e.g., Refs.[1–4] for reviews) greatly accelerate basis sets convergence of wavefunction ab initio methods. Experience from thermochemistry (e.g.[5–7]), noncovalent interactions (e.g.[8,9,10–12]) and vibrational frequencies[13,14] shows that explicitly correlated methods gain two,[15] or for larger basis sets even three, angular momentum increments on their conventional correlated counterparts. Because of the costs of coupled cluster methods scale steeply with the number of basis functions — which scales approximately as $\propto L^3$ with angular momentum L — this can mean a cost difference of an order of magnitude.

In current practice, Slater-type geminals,[16] better known as F12 geminals, have become the *de facto* standard for explicitly correlated methods. (For a recent comparison of alternative geminal forms, see Ref.[17])

CCSD(T), coupled cluster theory[18] with all singles and doubles[19] plus a quasiperturbative correction for triple excitations,[20,21] is widely considered the "gold standard" of ab initio quantum chemistry. The triples component does not benefit from F12 functions (despite attempts[22–24] to compensate by scaling; for attempts to corporate F12 into triples, see Refs.[25,26]) and hence we found[7,12] that it is preferable to obtain the triples contribution from conventional CCSD(T) calculations. At any rate, the (T) component is typically an order of magnitude smaller than the CCSD valence correlation, and higher-order correlation effects another order of magnitude smaller still (see the ESIs of Refs.[27,28] for tabulation for 140 and 200 molecules, respectively). Furthermore, (T) converges faster with the basis set[29–32] than CCSD, which leaves basis set convergence in the latter as the one 'accuracy bottleneck' where F12 has the most to offer.

Recently, however, claims have been proffered that CCSD and common explicitly correlated methods like CCSD-F12b do not converge to the same basis set limit. Such claims were made by Cremer and coworkers[33] for formic acid dimer, and by Feller[34] for total atomization energies of small molecules. While the formic acid discrepancy was ultimately ascribed to other sources,[35] it bears emphasizing that CCSD-F12b[36,37] and its more rigorous companion CCSD(F12*),[38] also known as CCSD-F12c in MOLPRO-speak, are *approximations* to the full CCSD-F12 method. From a purely formal viewpoint, the presence of the strong orthogonality projector guarantees that all F12 terms should eventually vanish as the one-electron basis becomes complete, and hence CCSD and all variants of CCSD-F12 should converge to the same complete basis set (CBS) limit. This does not guarantee, however, that this would be the case for finite basis sets small enough to be practically usable.

During our work on the W4-17 benchmark,[28] where the reference values were obtained by basis set extrapolation from conventional calculations, we also carried out explicitly correlated calculations using large basis sets for certain molecules. For many systems we found excellent agreement between both approaches, but for some others — notably those with significant nondynamical correlation[39] — we found discrepancies between CCSD and CCSD-F12b that appeared to persist even with large basis sets. Moreover, for those systems we found discrepancies *between different approximate F12 methods* that persisted even with large basis sets.

We then embarked on a comprehensive investigation, the results of which we report in the present paper.

## COMPUTATIONAL METHODS

<u>Choice of benchmark.</u> We chose the W4-17 benchmark[28] of 200 small first-and second-row molecules, which is an expanded update of the earlier W4-11 benchmark,[27] which in turn represents an expansion of the W4-08 benchmark[40] used for parametrizing the B2GP-PLYP DFT functional.[40] The W4-17 dataset is chemically diverse in that it spans a broad variety of bonding situations (covalent, ionic, strained, "hypervalent",...) as well as a broad range of electronic correlation character from predominantly dynamical correlation to pathological static correlation. The reference geometries were taken from the ESI of Ref.[28] and used without further modification. The W4-08 subset (96 molecules) will also be referred to in a few places in this paper.

To avoid clouding the issue, we focus exclusively on valence correlation in this paper. The basis set convergence of subvalence correlation for the W4-17 benchmark has very recently been studied in great detail;[41] it was found there that even conventional CCSD(T) extrapolated from aug-cc-pwCV{T,Q}Z basis sets[42] can reproduce this contribution to within 0.03 kcal/mol RMS, and CCSD(T)/aug-cc-pwCV{Q,5}Z to about 0.01 kcal/mol RMS. The use of core-valence F12 basis sets[43] for the CCSD component was also discussed in Ref.[41] (The comparatively fast convergence of this term owes much to the near-perfect cancellation of the core-core correlation energy, and the still substantial cancellation of the core-valence correlation energy, between the molecule and its proatoms.) Hence, this is not the accuracy-limiting factor in thermochemical studies: as pointed out repeatedly (e.g., Refs.[7,44]), that honor belongs to the valence correlation energy.

<u>Choice of basis sets.</u> For the explicitly correlated calculations, we considered three basis sets: the cc-pVTZ-F12 and cc-pVQZ-F12 basis set of Peterson and coworkers,[45] and the aug-cc-pwCV5Z basis set.[42] In a previous study,[24] we were able to show that this latter basis set yields results very close to reference data obtained with very large spdfgh uncontracted basis sets[46] (REF-h, in the notation of Ref.[24]). In addition, we considered the cc-pV5Z-F12 basis set[7,24] for a subset of molecules; as our available CPU time resources forced us to choose between either a complete set of aug-cc-pwCV5Z data (and partial cc-pV5Z-F12 results) or a complete cc-pV5Z-F12 data set and partial aug-cc-pwCV5Z, we decided to prioritize aug-cc-pwCV5Z, which is the larger and more complete of the two basis sets.

For the conventional calculations, we considered two families of basis sets. One is the correlation-consistent family aug-cc-pV(n+d)Z, n=T,Q,5,6,7, taken from Ref.[47] and references therein. The "+d" indicates that a tight d function is added[48] to second-row atoms to assist with the description of the 3d orbital, which in high oxidation states of these elements sinks low enough to become a back-donation acceptor.[49] As is our custom, we do not place diffuse functions on hydrogen, both in order to save CPU time and to enhance numerical stability. These basis sets are denoted by the shorthand haVnZ (heavy atom-augmented [correlation consistent] valence n-tuple zeta) basis set throughout the paper.

The second family are the nZaPa (uniformly convergent n-tuple zeta augmented polarized basis sets) of Petersson and coworkers,[32,50,51] which were designed for smooth uniform convergence of both SCF and valence correlation energies. (Unlike the correlation consistent family,[52] but akin to the Weigend-Ahlrichs basis sets,[53] the number of primitives in nZaPa varies from left to right in a row of the Periodic Table to ensure that the basis set incompleteness error stays approximately constant.) After some initial experimentation, we applied the same approximation for hydrogen as in haVnZ, namely that we are using the un-augmented basis sets on H. This is indicated by the notation nZaPha.

*Electronic structure codes.* Unless otherwise indicated, all conventional calculations were carried out using Gaussian 09 Rev.E01[54] and Gaussian 16 Rev.B01,[55] while all explicitly correlated calculations were carried out using TURBOMOLE 6.6 and 7.2.1.[56] For open-shell species, the Watts-Gauss-Bartlett version of ROCCSD was used throughout, i.e., orbitals were semicanonicalized before transformation (unlike in MOLPRO,[57] where semicanonicalization happens after integral transformation — see the Appendix to Ref.[58] for a brief explanation of the differences for open-shell cases). The options `int(nobasistransform,acc2e=13) iop(3/59=7,8/11=1)` were used in the Gaussian calculations.

For the F12 calculations, we used the RI-MP2 auxiliary basis sets of Refs.[59,60], and the JKFit basis sets of Ref.[61], together with the CABS (complementary auxiliary basis sets) of Ref.[62,63] have been employed in the cc-pVnZ-F12 calculations, while the aug-cc-pwCV5Z/MP2FIT basis set was used for both RI-MP2 and CABS in the aug-cc-pwCV5Z calculations.[60] As in our previous thermochemical papers, the geminal exponent was set to 1.4 $a_0^{-1}$ throughout. (In response to an early preprint of this work, the question arose whether the large differences for VTZ-F12 and VQZ-F12 were an artifact of our choice of geminal exponent, rather than the MP2-F12 optimized 'recommended' values of 1.0 for those basis sets. As shown in Figures S.1 and S.2 In the supporting information, recalculation does not qualitatively change our conclusion, and in fact the box plot ranges becomes *larger* at beta=1.0.)

For the larger basis sets, coupled cluster convergence problems were encountered in some instances: in most cases, these could be remedied by tightening the SCF convergence criterion and setting `$scftol=1.0E-16` (which tightens several associated parameters in the `ccsdf12` module of Turbomole) or even `$scftol=1.0E-18`.

*Basis set extrapolation.* For the nZaPa basis sets at the MP2 level, we used the extrapolation recommended by Ranasinghe and Petersson,[32] expressed here in the form of Schwenke: [30]

$$E_\infty = E(L) + A_L[E(L) - E(L-1)] \tag{1}$$

Here the Schwenke coefficient $A_L$ is specific to the basis set pair and level of electronic structure theory. For the other levels of theory we use a reparametrization of the same formula (Table 1). As shown in Ref.[64], this expression can be related quite simply to the more familiar extrapolation formulas:

$$E_L = E_\infty + \frac{B}{L^\alpha} \quad \text{if} \quad \alpha = \frac{\log\left(1 + \frac{1}{A_L}\right)}{\log\left(\frac{L}{L-1}\right)} \tag{2}$$

$$E_L = E_\infty + \frac{D}{(L+a)^3} \quad \text{if} \quad a = \frac{1}{\left(1 + \frac{1}{A_L}\right)^{1/3} - 1} + 1 - L \tag{3}$$

and conversely:

$$A_L = \frac{1}{\left(\frac{L+a}{L-1+a}\right)^\alpha - 1} \tag{4}$$

Petersson and coworkers observed repeatedly[32,51,65] that the CCSD–MP2 difference, at least for the nZaPa basis sets, asymptotically converges approximately as $(L+3/2)^{-3}$. Thus, CCSD-MP2 differences at the basis set limit can be obtained in two ways: using this formula, and as differences of extrapolated CCSD and MP2 CBS limit. Comparison offers one estimate of the residual uncertainty in these quantities.

Another estimate of the uncertainty derives from Eq. (1): the uncertainty in the extrapolation parameter times the RMS atomization energy difference between the two largest basis sets. As shown in Table 1, Schwenke coefficients between different basis set families such as nZaPa and haVnZ, or between different calibrations, are actually remarkably consistent. We believe that an uncertainty of ±0.1 in $A_L$ is actually a somewhat conservative estimate; for the 6,7}ZaPha and haV{6,7}Z+d basis set pairs, this would translate into a RMS uncertainty of about 0.06 kcal/mol at the MP2 level, and 0.04 kcal/mol at the CCSD level. For the CCSD-MP2 difference, this would be an even smaller 0.02 kcal/mol, but we would approximately double that amount on account of the greater uncertainty in the Schwenke parameter $A_L$. If we had limited ourselves to the {5,6} basis set pairs, all uncertainties would be approximately double those we have currently. For the discussion at hand, it means that the additional computational effort of the {6,7} basis set pairs is justified.

**TABLE 1.** Schwenke coefficients and equivalent Petersson shifts for different basis set pairs

|  | Basis sets | Schwenke coefficients $A_L$ | | | | Equivalent Petersson shifts $a$ | | | |
| --- | --- | --- | --- | --- | --- | --- | --- | --- | --- |
|  |  | {6,7} | {5,6} | {4,5} | {3,4} | {6,7} | {5,6} | {4,5} | {3,4} |
| $L^{-3}$ pure | generic | 1.701 | 1.374 | 1.049 | 0.730 | 0.00 | 0.00 | 0.00 | 0.00 |
| $(L+3/2)^{-3}$ pure | generic | 2.194 | 1.865 | 1.537 | 1.211 | 1.50 | 1.50 | 1.50 | 1.50 |
| Martin[a] | MP2/AVnZ | 1.852 | 1.503 | 1.127 |  | 0.46 | 0.40 | 0.24 |  |
| Petersson[32] opt. | MP2/nZaPa | 1.865 | 1.519 | 1.185 | 0.886 | 0.50 | 0.45 | 0.42 | 0.49 |
| Hill et al.[46] | MP2/AVnZ | N/A | 1.478 | 1.186 | 0.933 | N/A | 0.32 | 0.42 | 0.64 |
| Martin[b] | CCSD/AVnZ | 1.602 | 1.283 | 0.932 |  | -0.30 | -0.28 | -0.36 |  |
| Martin[b] | CCSD/nZaPa | 1.605 | 1.232 | 0.917 |  | -0.29 | -0.44 | -0.41 |  |
| Varandas[66,67] | CCSD/AVnZ | N/A | 1.295 | 0.912 | 0.665 | N/A | -0.24 | -0.43 | -0.21 |
| Schwenke[30] | CCSD/AVnZ | N/A | 1.266 | 0.930 | 0.700 | N/A | -0.33 | -0.37 | -0.09 |

(a) Ref.[64] fitted to MP2-F12/REF-h[46] data obtained using MOLPRO 2015.[57] Aux. basis sets from Ref.[46]
(b) Ref.[64] fitted to CCSD-F12 data for 12 closed-shell species in ESI of Ref.[13] at ref. geoms. *ibid.* Original aug-cc-pV7Z basis sets taken from Ref.[47] and refs. therein; updates courtesy of Dr. David Feller (PNNL).

## RESULTS AND DISCUSSION

### Quality and convergence of the conventional reference data

*MP2 convergence behavior using different approaches.* We were able to obtain MP2/haV{6,7}Z+d data for 188 out of the 200 species, and MP2/{6,7}ZaPha as well as MP2-F12/awCV5Z data for the complete set of 200. The root mean square difference (RMSD) between MP2-F12/awCV5Z and MP2/haV{6,7}Z+d is 0.095 kcal/mol, but the RMSD between MP2-F12/awCV5Z and MP2/{6,7}ZaPha is only 0.064 kcal/mol. For the W4-08 subset of molecules, both RMSDs are comparable: 0.067 and 0.065 kcal/mol, respectively. The main sources of the difference appear to be a number of chlorine compounds such as $CCl_4$ and $C_2Cl_6$, for which the haVnZ+d basis sets appear to exhibit oscillatory convergence.

As can be seen in Figure 1, this is reflected in surprisingly large errors for some of these systems even with basis sets as large as haV{5,6}Z+d. Also by way of illustration, the RMSD between MP2/{5,6}ZaPha and MP2/{6,7}ZaPha is 0.13 kcal/mol, but increases to 0.17 kcal/mol between MP2/haV{5,6}Z+d and MP2/haV{6,7}Z+d. We have previously[7] discussed the overcontraction issues of the haVnZ+d basis sets for second-row atoms. As further evidence, for the {5,6} basis set pair, the RMSD between extrapolated MP2 values with nZaPa and haVnZ+d basis sets is 0.20 kcal/mol, which drops to 0.07 kcal/mol for the {6,7} pair.

In Ref.[7] we observed that increasing the radial flexibility of the aug-cc-pV(5+d)Z and aug-cc-pV(6+d)Z basis sets, e.g. by adding core-valence basis functions of the lower angular momenta, greatly remediated the oscillatory convergence observed especially for chlorine compounds. We have taken the "maximalist" of the different approaches discussed there and added *spd* core-valence functions from the corresponding aug-cc-pCVnZ basis set — this combination we have denoted haV5Z+C(spd). (As the core-valence d functions already cover the tight d required for 3d orbital back-bonding at the SCF level, the "+d" addition was not required.) As can be seen in Figure 1, this does bring some succor, but not as much as a large basis set MP2-F12 calculation.

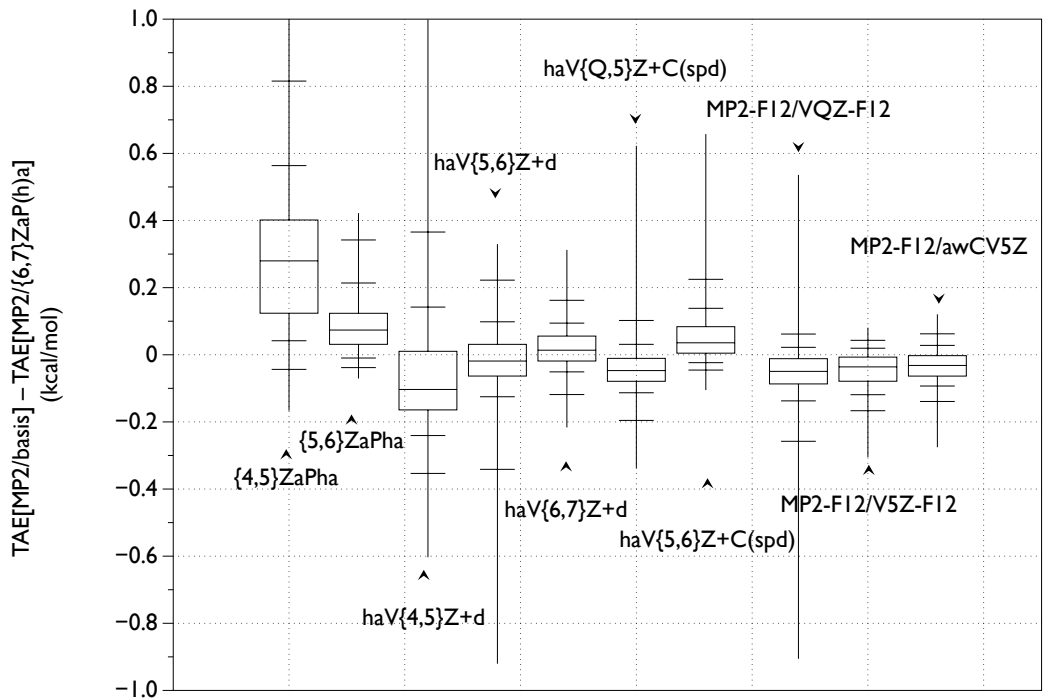

**FIGURE 1.** Box plot of deviations from MP2/{6,7}ZaP(h)a reference values (kcal/mol) for the valence MP2 correlation components of the W4-17 atomization energy benchmark. The outer fences encompass the middle 95% of the distribution, the inner fences 80%, the box 50%. Vertical lines span from population minimum to maximum.

<u>CCSD-MP2 differences considered.</u> We were able to obtain CCSD/{6,7}ZaPha values for 192 out of 200 molecules, and CCSD(F12*)/awCV5Z values for all species except $C_6H_6$. Comparison of individual differences between MP2 and MP2-F12 for molecules that ought not present convergence issues with the {6,7} pair, such as alkanes, suggests that the hydrogen basis set in MP2-F12/awCV5Z may still be insufficiently saturated, although adequate saturation does appear to have been achieved for nonhydrogen species. We do not, however, expect the CCSD–MP2 difference to be appreciably affected by this, as these small discrepancies (about 0.01 kcal/mol per hydrogen, cf. Ref.[24]) can reasonably be expected to cancel between CCSD and MP2. For instance, for propane, the MP2-F12/awCV5Z correlation contribution of 224.42 kcal/mol is more than 0.1 kcal/mol below the MP2/haV{6,7}Z and MP2/{6,7}ZaPha values of 224.56 kcal/mol and 224.53 kcal/mol, respectively. Yet for the CCSD(F12*)–MP2-F12 difference with the awCV5Z basis set, we obtain –15.61 kcal, which is considerably closer to the conventional {6,7}ZaPha values of –15.66 kcal/mol computed as CCSD/CBS – MP2/CBS, and –15.62 kcal/mol using Petersson's $(L+3/2)^{-3}$ formula applied to the CCSD-MP2 differences of the {6,7} pair. The corresponding conventional values for the haV{6,7}Z pair are –15.68 and –15.66 kcal/mol, respectively.

Hence, we have compared the CCSD–MP2 differences between {6,7}ZaPha and haV{6,7}Z+d, and found the two sets of data to differ by just 0.034 kcal/mol RMSD. In addition, {5,6}ZaPha and {6,7}ZaPha differ by just 0.06 kcal/mol RMSD, and a similarly small RMSD of 0.07 kcal/mol is seen between haV{5,6}Z+d and haV{6,7}Z+d. This bolsters our confidence in the quality of the CCSD-MP2 differences with the {6,7} basis set pairs.

Can we legitimately omit the diffuse functions on H in {6,7}ZaPha? Particularly for the Al and Si compounds, placing no diffuse functions on H while attaching them to the metal and metalloid, respectively, seems dubious, and even for B this choice can be called into question. For the W4-08 subset, we performed calculations using {6,7}ZaPa in which diffuse functions were *not* omitted on hydrogen. MP2/{6,7} differences between the {6,7}ZaPha and {6,7}ZaPa basis set pairs were found to be 0.01 kcal/mol or less for the HBCNOFPSCl compounds, but reached 0.05 kcal/mol for $Si_2H_6$, 0.04 kcal/mol for $SiH_4$, and 0.03 kcal/mol for $AlH_3$. The only Al or Si hydride in W4-17 that is not an element of the W4-08 subset is $SiH_3F$, for which we found a difference of 0.03 kcal/mol upon recalculation.

We have hence selected, as our reference level, {6,7}ZaPa for SiH$_3$F and the W4-08 subset, and {6,7}ZaPha for the remainder. This also circumvents the oscillatory convergence issues we noted for the haVnZ+d sequence.

The CCSD-MP2 values were obtained by taking the differences between extrapolated CCSD and MP2 limits. In principle, we could also consider the CCSD-MP2 differences directly, and extrapolate them according to the $(L+3/2)^{-3}$ formula observed empirically by Petersson and coworkers.[65,68] For the {6,7} pair, the RMSD from the difference-of-limits values is 0.05 kcal/mol for the nZaPha basis sets.

We also investigated the use of "interference corrections",[69] but the results were erratic, and we have not retained them for our analysis.

Based on the RMS differences between the {6,7}ZaPha and haV{6,7}Z+d data, as well as between extrapolation of differences and differences of extrapolations, we would estimate the remaining uncertainty of our chosen reference level, i.e. [CCSD-MP2]/{6,7}ZaPha, to be on the order of 0.03–0.04 kcal/mol.

## Agreement with approximate CCSD-F12 data near the basis set limit

We are now in a position to compare our best CCSD-MP2 values with the CCSD-F12 – MP2-F12 differences obtained using various approximate CCSD-F12 methods. The aug-cc-pwCV5Z basis set was employed throughout.

First, we can see in Table 2 and Figure 2 that CCSD(F12*) has an RMSD of only 0.035 kcal/mol, comparable to the uncertainty in the reference values.

**FIGURE 2.** Box plot of deviations from our best estimates for CCSD-MP2 components of W4-17 atomization energies (kcal/mol) using the awCV5Z basis set and different approximations to CCSD-F12

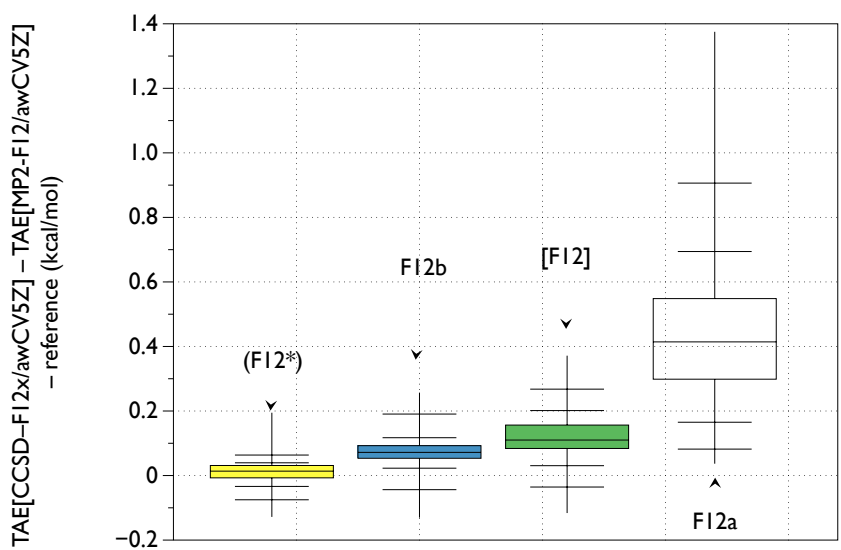

CCSD(F12*) is itself an approximation to the more rigorous (and much more costly) CCSD(F12) method.[70] It has been shown before[38] that the difference between them is very small, but it does bear verifying this for the present W4-17 atomization energy benchmark. For the cc-pVTZ-F12 basis set, the difference between CCSD(F12*) and CCSD(F12) atomization energies is just 0.01 kcal/mol RMS, reaching a maximum of 0.035 kcal/mol for HClO$_4$. These differences (Figure 3) become negligible altogether when expanding the basis set to cc-pVQZ-F12: RMS 0.002 kcal/mol, with maximum of 0.008 kcal/mol for C$_2$Cl$_6$. For the W4-08 subset we finally considered the cc-pV5Z-F12 dataset and found differences of less than 0.001 kcal/mol RMS that should be considered thermochemically equivalent within the numerical noise of the calculation. We conclude that for cc-pVQZ-F12 or larger basis sets, CCSD(F12*) is functionally equivalent to CCSD(F12).

While we cannot definitely rule out that post-CCSD(F12) corrections might be nontrivial for *some* molecules, verifying this is not technically feasible at present, both because of code limitations and because the changes would be below the resolution level of our reference data. (A few examples have been investigated in Ref.[71] using the cc-

pVTZ-F12 basis set. It was found that while the individual contributions from terms beyond CCSD(F12) can be sizable, they cancel very systematically.)

**TABLE 2.** RMSD (kcal/mol) for the W4-17 TAE Benchmark from our best CCSD/{6,7}ZaP(h)a – MP2/{6,7}ZaP(h)a limits for different CCSD-F12 approximations. The aug-cc-pwCV5Z basis set with beta=1.4 was used throughout the F12 calculations.

|         | CCSD(F12*)–MP2-F12 | CCSD-F12b–MP2-F12 | CCSD[F12]–MP2-F12 | CCSD-F12a–MP2-F12 |
|---------|--------------------|-------------------|-------------------|-------------------|
| RMSD    | 0.035              | 0.087             | 0.137             | 0.470             |
| MSD     | 0.008              | 0.071             | 0.116             | 0.424             |
| StDev[a]| 0.034              | 0.050             | 0.073             | 0.202             |

(a) calculated as $(RMSD^2 - MSD^2)^{1/2}$

**FIGURE 3.** Box plot of basis set convergence of CCSD(F12) – CCSD(F12*) differences (kcal/mol) for the W4-17 dataset

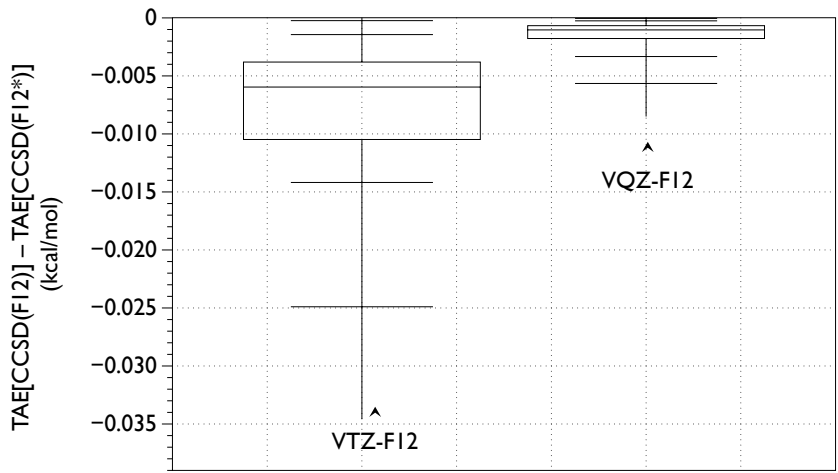

In contrast, for CCSD-F12b we see an RMSD that is about 2.5 times as large and cannot easily be blamed on uncertainty in the reference values anymore. (In fact, 0.08 kcal/mol is about the RMSD of W4 theory for its training set.) This result is consistent with that of earlier, less rigorous, benchmark studies on smaller samples and using smaller basis sets.[38,72] CCSD-F12b does represent a dramatic improvement over the still more approximate CCSD-F12a method, which has an error of 0.43 kcal/mol RMS even with such large basis sets.

Köhn and Tew[71] carried out a detailed multiple perturbation theory analysis of approximations to CCSD-F12 in terms of five coupling strengths : single ($\sigma$) and double ($\lambda$) excitation amplitudes into the virtual orbitals, single ($\tau$) and double ($\mu$) excitations into the auxiliary basis set, and finally $\nu$, which corresponds to mixed double excitations (one electron into virtual, the other into auxiliary space). The key points of the analysis are summarized in Table 3: for mathematical details, we refer to Ref.[71] itself, specifically Tables I, II, and VII and the surrounding discussion.

The multiple perturbation expansion is winnowed by restricting the total order in $\mu$ and $\nu$ to at most two. Retaining then all remaining terms through third order corresponds to the CCSD[F12] method,[71] which was not recommended for practical use (by its authors) but turns out to be useful in our analysis. The CCSD(F12*) method is obtained (see Table VII of Ref.[71]) by adding a group of fourth-order terms quadratic in $\mu$, namely, of orders $\sigma^2\mu^2$, $\sigma\lambda\mu^2$, and $\lambda^2\mu^2$, plus a single fifth-order term of order $\sigma^2\lambda\mu^2$. In contrast, beyond second order, CCSD-F12b[36,37] and Valeev's CCSD(2)$_{\overline{F12}}$ method[73] only contain the third-order contributions of orders $\sigma\mu^2$ and $\lambda\mu^2$, plus fourth-order contributions scaling as $\sigma^2\mu^2$ and $\sigma\lambda\mu^2$. Finally, CCSD-F12a discards all CCSD-F12b contributions higher than second order to the energy equation, but not in the amplitude equations.

**TABLE 3.** Terms included in CCSD(F12) and approximations thereto. For further details, see Tables I and II in Ref.[71] and the surrounding discussion

| Parent expression | Label | Order | (F12) | (F12*) | [F12] | F12b | F12a | $(2)_{\overline{F12}}$ | MP2-F12+ΔCCSD |
|---|---|---|---|---|---|---|---|---|---|
| $\langle 0\lvert G\tilde{R} + \tilde{R}^\dagger G + \tilde{R}^\dagger F\tilde{R}\rvert 0\rangle$ | $E_{\text{MP2}}^{\text{unc.}}$ | $\mu^2$ | ✓ | ✓ | ✓ | ✓ | ✓ | ✓ | ✓ |
| $\langle 0\lvert \tilde{R}^\dagger F^{(0)}T_2\rvert 0\rangle$ | R.1 | $\lambda\nu$ | ✓ | ✓ | ✓ | ✓ | ✓ | | |
| $\langle 0\lvert \tilde{R}^\dagger[G,T_2]\rvert 0\rangle$ | R.2 | $\lambda\mu^2$ | ✓ | ✓ | ✓ | ✓ | | ✓ | |
| | R.3 | $\lambda\nu^2$ | ✓ | ✓ | ✓ | | | | |
| $\langle 0\lvert \tilde{R}^\dagger[G,T_1]\rvert 0\rangle$ | R.4 | $\sigma\mu^2$ | ✓ | ✓ | ✓ | ✓ | | ✓ | |
| | R.5 | $\sigma\nu^2$ | ✓ | ✓ | ✓ | | | | |
| $\langle 0\lvert \tilde{R}^\dagger[[G,T_1],T_2]\rvert 0\rangle$ | R.6 | $\sigma\lambda\nu^2$ | ✓ | | | | | | |
| $\frac{1}{2}\langle 0\lvert \tilde{R}^\dagger[[G,T_1],T_1]\rvert 0\rangle$ | R.7 | $\sigma^2\mu^2$ | ✓ | ✓ | | ✓ | | | |
| | R.8 | $\sigma^2\nu^2$ | ✓ | | | | | | |
| $\frac{1}{6}\langle 0\lvert \tilde{R}^\dagger[[[G,T_1],T_1],T_1]\rvert 0\rangle$ | R.9 | $\sigma^3\nu^2$ | ✓ | | | | | | |
| $\langle 0\lvert \Lambda_1[G,\tilde{R}]\rvert 0\rangle$ | L1.1 | $\sigma\mu^2$ | ✓ | ✓ | ✓ | ✓ | ✓ | ✓ | |
| | L1.2 | $\sigma\nu^2$ | ✓ | ✓ | ✓ | | | | |
| $\langle 0\lvert \Lambda_1[F^{(1)},\tilde{R}]\rvert 0\rangle$ | L1.3 | $\sigma\tau\nu$ | ✓ | ✓ | ✓ | | | | |
| $\langle 0\lvert \Lambda_1[[G,T_1],\tilde{R}]\rvert 0\rangle$ | L1.4 | $\sigma^2\mu^2$ | ✓ | ✓ | | | | | |
| | L1.5 | $\sigma^2\nu^2$ | ✓ | | | | | | |
| $\langle 0\lvert \Lambda_2 F^{(0)}\tilde{R}\rvert 0\rangle$ | L2.1 | $\lambda\nu$ | ✓ | ✓ | ✓ | ✓ | ✓ | | |
| $\langle 0\lvert \Lambda_2[G,\tilde{R}]\rvert 0\rangle$ | L2.2 | $\lambda\mu^2$ | ✓ | ✓ | ✓ | ✓ | ✓ | ✓ | |
| | L2.3 | $\lambda\nu^2$ | ✓ | ✓ | ✓ | | | | |
| $\langle 0\lvert \Lambda_2[[G,T_2],\tilde{R}]\rvert 0\rangle$ | L2.4 | $\lambda^2\mu^2$ | ✓ | ✓ | | | | | |
| | L2.5 | $\lambda^2\nu^2$ | ✓ | | | | | | |
| $\langle 0\lvert \Lambda_2[[G,T_1],\tilde{R}]\rvert 0\rangle$ | L2.6 | $\sigma\lambda\mu^2$ | ✓ | ✓ | | ✓ | ✓ | | |
| | L2.7 | $\sigma\lambda\nu^2$ | ✓ | | | | | | |
| $\langle 0\lvert \Lambda_2[[F^{(1)},T_1],\tilde{R}]\rvert 0\rangle$ | L2.8 | $\sigma\tau\lambda\nu$ | ✓ | | | | | | |
| $\frac{1}{2}\langle 0\lvert \Lambda_2[[[G,T_1],T_1],\tilde{R}]\rvert 0\rangle$ | L2.9 | $\sigma^2\lambda\mu^2$ | ✓ | ✓ | | | | | |
| | L2.10 | $\sigma^2\lambda\nu^2$ | ✓ | | | | | | |

G, $T_1$, $T_2$, have their usual respective meanings of the two-electron repulsion operator, the single substitutions operator, and the double substitutions operator. The matrix elements of $\tilde{R}$ are those of the geminal function, while $\Lambda_1$ and $\Lambda_2$ are de-excitation operators associated with the Lagrange multipliers for the singles and doubles coupled-cluster residual equation the singles and doubles Lagrange multiplier terms. F(0) is the Fock operator less those terms that are only non-zero if the Brillouin or the generalized Brillouin theorem are not fulfilled, which are summarized as F(1). For further details, see Ref.[71]

In the present work, we found CCSD[F12] to have an RMSD slightly higher than CCSD-F12b, 0.11 kcal/mol. Some additional light is shed by the mean signed difference (MSD), as well as by the standard deviation about it (StDev). While CCSD(F12*) pleasingly has an RMSD close to zero (indicating an absence of systematic bias), CCSD-F12b, CCSD[F12], and CCSD-F12a have progressively larger systematic biases. In fact, the StDev values of CCSD-F12b and CCSD[F12] are somewhat comparable, while that for CCSD-F12a is about three times as large. The statistical behaviors of the various approximations are depicted as a box plot in Figure 2.

## Basis set convergence of differences between various F12 methods

Figure 4 illustrates the basis set convergence of the CCSD-F12b – CCSD-F12a difference. Said difference corresponds, in Table VII of Ref.[71], to the terms R.2($\lambda \cdot \mu^2$) + R.4($\sigma \mu^2$) + R.7 ($\sigma^2 \mu^2$), where the orders in perturbation theory are indicated in parentheses as powers of the coupling strengths.

**FIGURE 4.** Box plot of the CCSD-F12b – CCSD-F12a difference (kcal/mol) for the W4-17 dataset and different basis sets

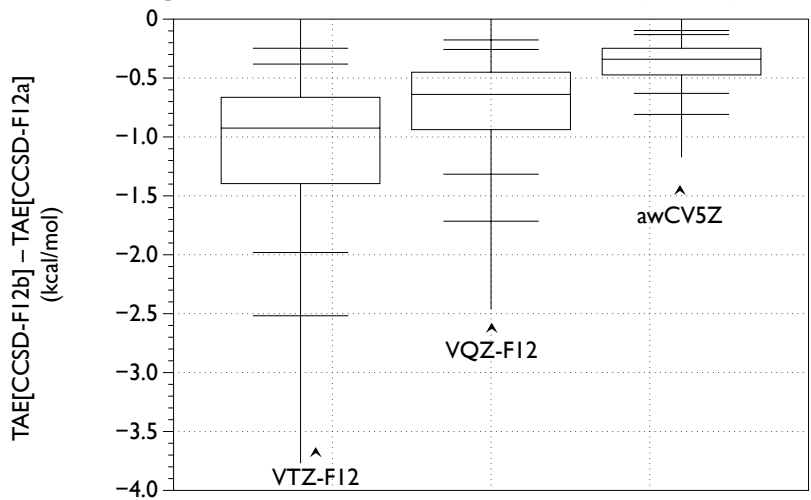

For cc-pVTZ-F12, which is the smallest basis set one would normally consider for thermochemical production applications, the median difference is close to 1 kcal/mol, and for 2.5% of our test species it actually exceeds 2.5 kcal/mol. For cc-pVQZ-F12, which is the largest basis set used in typical CCSD-F12x applications, the median F12b–F12a difference is still close to 0.7 kcal/mol, and values in excess of 1.5 kcal/mol can still be found. These numbers can be cut in half with awCV5Z, but this is hardly an acceptable error level for such large (by F12 standards) basis sets. We conclude that CCSD-F12a cannot be recommended at all for production work, although for small basis sets the results may seem superior to CCSD-F12b due to an error compensation[38] between the missing terms (which are antibonding) and basis set incompleteness error (see also Ref.[12] for an example and Refs.[3,4] for further discussion).

**FIGURE 5.** Box plot of the CCSD[F12] – CCSD-F12b difference (kcal/mol) for the W4-17 dataset and different basis sets

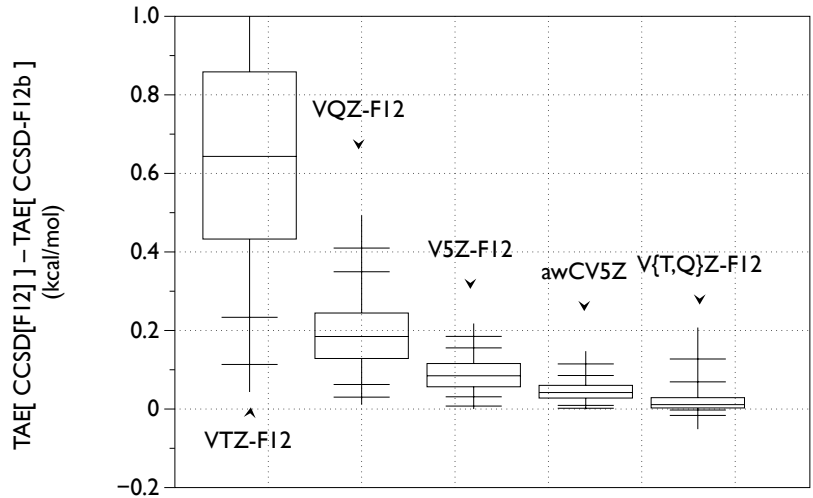

It turns out to be enlightening to decompose the CCSD(F12*) – CCSD-F12b difference into two components: CCSD(F12*) – CCSD[F12] and CCSD[F12] – CCSD-F12b. Energetically speaking (see Figures 3 and 4 in Ref.[71]), the most important third-order contributions are (following the notation in Table VII of Ref.[71]) the ladder terms R.2($\lambda.\mu^2$) and L2.2($\lambda.\mu^2$), followed by the ring terms R.3($\lambda.v^2$) and L2.3($\lambda.v^2$). (In this notation, R, L1, and L2 refer to the energy, singles amplitudes, and doubles amplitudes equations, respectively.) R.2 and L2.2 are already present in CCSD-F12b (in fact, L2.2 even in CCSD-F12a!), while CCSD[F12] adds in R.3($\lambda.v^2$) and L2.3($\lambda.v^2$), besides the smaller terms R.5 ($\sigma v^2$), L1.2 ($\sigma v^2$), and L1.3 ($\sigma \tau v$). The fourth-order terms R.7 ($\sigma^2\mu^2$) and L2.6 ($\sigma\lambda\mu^2$) from CCSD-F12b are omitted.

As can be seen in Figure 5, the CCSD[F12]–CCSD-F12b difference (a) is universally attractive; (b) decays very rapidly with the basis set. As the F12 corrections should in principle vanish in the complete basis set limit, basis set extrapolation should yield results close to zero. Such extrapolation from V{T,Q}Z-F12 using the formula of Hill et al.[74] shows a very narrow box with a median close to zero (Figure 4, far right): while there are clearly some outliers, the extrapolated values on the whole are somewhere between V5Z-F12 and awCV5Z in quality. This suggests that the convergence behavior is pretty regular even with fairly modest basis sets. (There are solid theoretical grounds to believe each ring term converges rapidly. For atoms, the partial wave expansion of terms with a single contraction over the CABS space was shown to be rapidly converging by Noga and Kutzelnigg (appendix C of Ref.[75]). This is the reason why they are neglected in the Standard Approximation.)

While it might be remotely possible that, for large basis sets, the seven terms in the CCSD[F12]–CCSD-F12b difference might simply benefit from an unusually felicitous form of mutual error cancellation, the Occam's Razor explanation for the rapid tapering off of their sum with increasing basis set would seem to be that the individual terms converge fairly rapidly.

Basis set convergence of the CCSD(F12*)–CCSD[F12] difference is depicted in Figure 6. As can be seen there, these contributions are universally repulsive and do taper off with the basis set: median values are about halved with each successive basis set, from about 0.4 kcal/mol for cc-pVTZ-F12 to 0.2 kcal/mol for cc-pVQZ-F12 to 0.1 kcal/mol for awCV5Z. Crucially, however — even for basis sets as large as awCV5Z — they remain nontrivial (0.1 kcal/mol median), exceeding 0.2 kcal/mol for about 10% of the sample and 0.3 kcal/mol for about 2.5%. What's more, the distribution is strongly skewed/asymmetric. Furthermore, extrapolation does not help for this term: the error distribution of V{T,Q}Z-F12 looks no better than that of VQZ-F12 itself, and likewise for V{Q,5}Z-F12 using the extrapolation from Table 2 in Ref.[11]

In the notation of Ref.[71], the difference corresponds to the following fourth-order terms: R.7 ($\sigma^2\mu^2$) in the energy expression, L1.4 ($\sigma^2\mu^2$) in the singles amplitudes equation, and **L2.4 ($\lambda^2\mu^2$)** + L2.6 ($\sigma\lambda\mu^2$) + L2.9 ($\sigma^2\lambda\mu^2$) in the doubles amplitudes equation. Out of these, R.7 and L2.6 correspond to restoring the two F12b terms deleted in CCSD[F12] (which can hence be assumed to be small), which leaves L1.4, L2.4, and L2.6. For a much smaller sample of molecules and just the cc-pVDZ-F12 and cc-pVTZ-F12 basis sets, Figure 3 in Ref.[71] shows a plot of the contribution of the various terms to the correlation energy (normalized by the number of valence electrons): L2.4 is by far the most important, followed by L1.4 two orders of magnitude lower and L2.6 another order of magnitude below.

**FIGURE 6.** Box plot of the CCSD(F12*) – CCSD[F12] difference (kcal/mol) for the W4-17 dataset and different basis sets

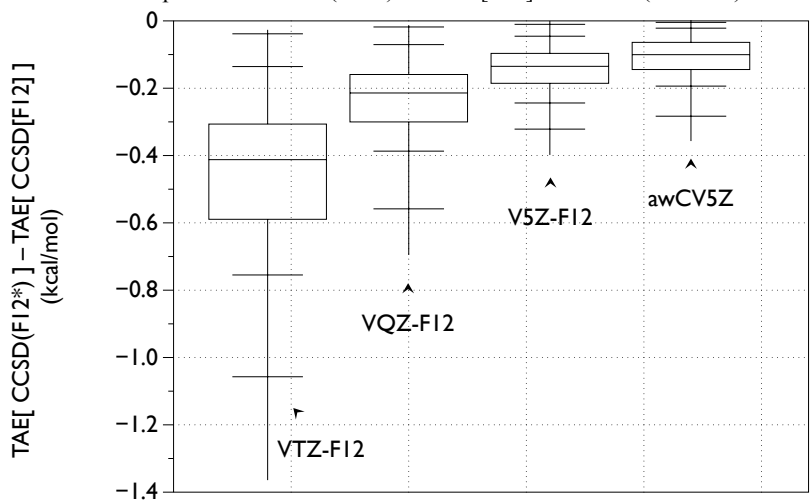

Detailed inspection of the molecules for which the CCSD(F12*)–CCSD[F12] difference is large reveals an intriguing pattern: among small species, notorious multireference cases like $C_2(1\Sigma^+_g)$, $BN(1\Sigma^+)$, $O_3$, $FO_2$, $ClOO$, immediately jump out. If we normalize the difference by the number of valence electrons, then all the values in excess of 10 cal/mol/e$^-$ are seen for molecules with significant static correlation, such as $N_2O_4$, $HO_3$, $FO_2$, $F_2O_2$, $P_4$, $S_4$, $ClOO$, $NO_2$, $N_2O$, $O_3$, $BN(1\Sigma^+)$, $C_2(1\Sigma^+_g)$, $B_2(3\Sigma^-_g)$, $P_2$, and $ClO$.

L2.4 is a first-order geminal correction to a disconnected quadruples ($\hat{T}_2^2/2$) term, as is L1.4 to the disconnected doubles $\hat{T}_1^2/2$. It stands to reason that in situations where some doubles amplitudes $T_2$ are large, L2.4 will become important, and to a lesser extent, so will L1.4 if some singles amplitudes $T_1$ are large. (If some of both are large, the $T_1T_2$-geminals coupling term L2.6 could become significant.). Such scenarios occur, of course, when there is significant static correlation in the molecule.

Coming back to the CCSD(F12*)–CCSD-F12b difference, the superposition of two terms with opposite sign and different (fast vs. relatively slow) convergence rates leads to the convergence behavior seen in Figure 7.

**FIGURE 7.** Box plot of the CCSD(F12*) – CCSD-F12b difference (kcal/mol) for the W4-17 dataset and different basis sets

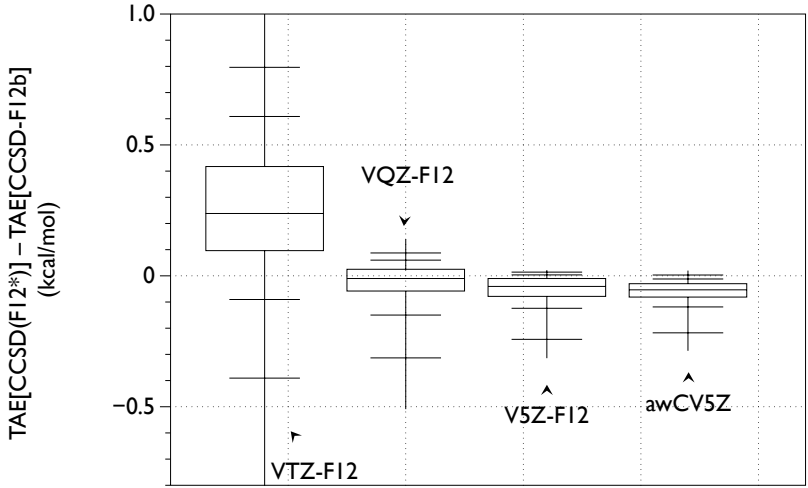

The reader may wonder how these various terms evolve as a molecule grows, e.g. along the n-alkane sequence $CH_3(CH_2)_nCH_3$ (n=0,1,2,3) or the sequence ethylene, trans-butadiene, and 1,3,5-hexatriene. As seen in Table 4 for the cc-pVTZ-F12 and cc-pVQZ-F12 basis sets, the [F12]–F12b and (F12*)–[F12] differences grow almost perfectly linearly along the sequences, as expected for strictly size extensive methods like coupled-cluster theory. In turn, size extensivity means that all conclusions from this paper carry over to arbitrarily large systems, as long as strong long-range correlations do not play a role (in which case single-reference CCSD breaks down anyway).

**TABLE 4.** Chain length dependence of differences (kcal/mol) between different CCSD(F12) approximations.

|  | VTZ-F12 | | | VQZ-F12 | | |
|---|---|---|---|---|---|---|
|  | [F12]-F12b | (F12*)–[F12] | (F12*)-F12b | [F12]-F12b | (F12*)–[F12] | (F12*)-F12b |
| n-pentane | 1.597 | -0.581 | 1.016 | 0.441 | -0.300 | 0.140 |
| n-butane | 1.293 | -0.467 | 0.826 | 0.356 | -0.242 | 0.114 |
| propane | 0.988 | -0.353 | 0.634 | 0.270 | -0.183 | 0.087 |
| ethane | 0.682 | -0.239 | 0.443 | 0.184 | -0.124 | 0.060 |
| methane | 0.372 | -0.127 | 0.245 | 0.096 | -0.066 | 0.030 |
| hexatriene | 1.653 | -0.706 | 0.947 | 0.455 | -0.368 | 0.088 |
| butadiene | 1.123 | -0.475 | 0.648 | 0.308 | -0.247 | 0.060 |
| ethylene | 0.593 | -0.244 | 0.349 | 0.160 | -0.127 | 0.033 |

Finally, let us address the basis set convergence of the most rigorous approximation to CCSD-MP2, i.e., CCSD(F12*)–MP2-F12. This is depicted in Figure 8. As can be seen there, one needs at least a cc-pVQZ-F12 basis set for a satisfactory error distribution.

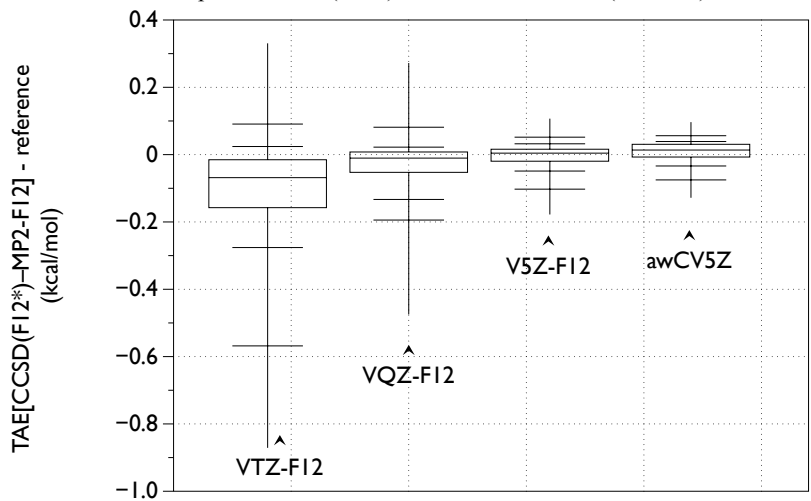

**FIGURE 8.** Box plot of CCSD(F12*)–MP2-F12 difference (kcal/mol) for the W4-17 dataset and different basis sets

## Relationship to static correlation diagnostics

It would, of course, be helpful to have an *a priori* prediction for whether the (F12*)–F12b difference is nontrivial. Thus, establishing a link with static correlation character would serve a pragmatic as well as an interpretative purpose.

Figure 9 depicts the Pearson correlation coefficients R between various static correlation diagnostics and the differences between various F12 approximations. In this table, the cc-pVQZ-F12 basis set was used for the F12 results, as the (F12*)-[F12] term, in particular, was deemed too small with the awCV5Z basis set to yield useful conclusions.

It should be kept in mind that the various diagnostics do not all measure the same thing. Hollett and Gill[39] (see also Scuseria and Tsuchimochi[76]) distinguish between "type A static correlation" (absolute near-degeneracy, with dissociating $H_2$ as the paradigmatic example) and "type B static correlation" (relative near-degeneracy, with Be-like ions $(Z)^{(Z-4)+}$ as a textbook case). The von Neumann correlation entropy $S_{corr}$, for instance (a.k.a., "entanglement entropy"), is strongly linked to type A static correlation, while pragmatic diagnostics like the percentage of connected triples in the total atomization energy, %TAE[(T)],[58] or the percentage of post-CCSD(T) correction are primarily concerned with thermochemical importance. (Note that these latter diagnostics remain identically zero for the Be-like ions, as CCSD is an exact solution for the two-electron problem.) The DFT-based $A_{25}$[PBE] diagnostic,[77] which is based on the slope of the DFT atomization energy as a function of the HF-like exchange percentage, likewise is more pragmatic in character, although it has been argued[77] that it primarily samples type B static correlation.

Very recently, Matito and coworkers[78] proposed two new diagnostics for the importance of nondynamical and dynamical correlation that are based on the natural orbital occupations. While $I_{ND}$ and $I_D$ are not intensive but extensive (their values for a dimer at infinite separation are exactly the sums of the respective monomer values), the quantity $I_{ND}/(I_{ND}+I_D)$ can be considered as an intensive quantity (like the popular $T_1$ diagnostic[79], which is the Euclidian norm of the single substitutions vector divided by the square root of the number of correlated electrons).

While none of the correlations in Table 3 are good enough to permit quantitative estimation by linear regression, we *can* identify a few correlations of 0.8 or better. In particular, the "pragmatic" thermochemical diagnostics %TAE[(T)] and $A_{25}$[PBE] have correlation coefficients of about –0.8 with the (F12*)–F12b difference.

(%TAE[$T_4+T_5$], unless needed anyway as part of a W4, HEAT, or FPD calculation, is simply too costly to serve as an *a priori* estimate.) Matito's diagnostic and the correlation entropy $S_{corr}$, in the other hand, have equally good positive correlations with the [F12]–F12b difference. (We note that for the W4-17 dataset, we find R=0.994 between Matito's $I_{ND}$ and $S_{corr}$, indicating that they largely tell the same story.) If, on the other hand, we normalize the (F12*)–F12b difference by the number of valence electrons, we get a negative correlation R=–0.83 with the $T_1$ diagnostic.

**Figure 9.** Pearson correlation coefficients R between F12x differences and various nondynamical correlation diagnostics. In the red-white-blue color 'heatmap', red refers to large positive R, blue to large negative R, and white to near-zero R.

|  | unnormalized | | | | Normalized by number of valence electrons | | | |
|---|---|---|---|---|---|---|---|---|
|  | F12b-F12a | (F12*)-F12b | [F12]-F12b | (F12*)-[F12] | F12b-F12a | (F12*)-F12b | [F12]-F12b | (F12*)-[F12] |
| %TAE[(T)] | 0.125 | **-0.796** | 0.001 | -0.631 | 0.308 | -0.761 | -0.205 | -0.664 |
| %TAE[post-(T)] | 0.248 | -0.632 | -0.226 | -0.400 | 0.279 | -0.614 | -0.293 | -0.489 |
| %TAE[$T_4+T_5$] | 0.112 | **-0.785** | -0.037 | -0.597 | 0.222 | **-0.797** | -0.150 | -0.707 |
| $A_{25}$[PBE] | 0.191 | **-0.776** | -0.074 | -0.590 | 0.353 | -0.721 | -0.282 | -0.615 |
| $T_1$ diagnostic | 0.059 | -0.644 | -0.059 | -0.463 | 0.020 | **-0.834** | 0.013 | -0.768 |
| $D_1$ diagnostic | 0.004 | -0.721 | 0.043 | -0.573 | 0.090 | **-0.799** | 0.003 | -0.749 |
| Truhlar $M_{diag}$ | 0.058 | -0.562 | -0.057 | -0.394 | 0.026 | -0.729 | -0.031 | -0.640 |
| Matito $I_{ND}$ | -0.510 | -0.524 | **0.792** | -0.726 | 0.209 | -0.171 | 0.160 | -0.207 |
| Matito $I_D$ | -0.539 | -0.374 | **0.841** | -0.643 | 0.257 | 0.059 | 0.123 | -0.003 |
| $D_2$ diagnostic | -0.013 | **-0.780** | 0.104 | -0.652 | 0.132 | **-0.842** | **-0.826** | **-0.826** |
| $S_{corr}$ | -0.525 | -0.460 | **0.815** | -0.690 | 0.225 | -0.075 | -0.122 | -0.122 |
| $r_{ND}=I_{ND}/(I_{ND}+I_D)$ | 0.093 | -0.369 | -0.113 | -0.214 | -0.061 | -0.640 | 0.027 | -0.564 |
| $r_{ND} E_{corr}$(CCSD) | 0.341 | 0.623 | -0.683 | **0.804** | -0.395 | 0.248 | 0.023 | 0.270 |

It should be clarified that the minus sign in front of the larger negative correlations reflects negative (antibonding) contributions to the atomization energy. For total energies, the signs would be reversed.

The $D_1$[80] and especially $D_2$ diagnostics[81] also bear mentioning. While $T_1$ corresponds to a vector norm divided by the square root of the number of valence electrons, $D_1$ corresponds to a matrix norm, i.e. the square root of the largest eigenvalue of $T_1 \cdot T_1^\dagger$. Similarly, $D_2$ is obtained as the root of the largest eigenvalue of the double excitations amplitude matrix multiplied by its transpose, $T_2 \cdot T_2^\dagger$. Because of the mathematical properties of matrix norms, $D_1$ and $D_2$ are much more likely to reflect the part of the molecule with the most severe static correlation. (This becomes apparent from considering the didactic example of BN…n-octane at long separation. The large $T_1$ diagnostic of BN will be 'diluted' by the small one of the alkane, while $D_1$ and $D_2$ are identical to those of BN monomer.) It would seem to be reasonable that a diagnostic for important double excitation amplitudes would be at least a semiquantitative predictor for CCSD(F12*)–CCSD-F12b differences, considering that they appear to be driven by the L2.4 double excitations coupling term.

We note also in the full correlation matrix between the diagnostics (see Table S1 in the ESI) that $D_2$ correlates considerably better with the energetic diagnostics like %TAE[(T)] than does $T_1$ or $D_2$.

## PERSPECTIVE AND CONCLUSIONS

Following an extensive survey for a large thermochemical benchmark, we are in a position to conclude the following:

(a) Near the one-particle basis set limit, the difference between CCSD(F12*) and conventional CCSD/CBS is comparable to the uncertainty in the latter (less than 0.04 kcal/mol RMS)
(b) The difference between CCSD(F12*) and CCSD(F12) becomes thermochemically negligible (0.002 kcal/mol RMS) already with the cc-pVQZ-F12 basis set.
(c) We hence have no indication that CCSD-F12 terms beyond CCSD(F12) would play a significant thermochemical role with large basis sets
(d) In contrast, a thermochemically significant difference between CCSD(F12*) and the more widely used CCSD-F12b approximation remains even with basis sets as large as aug-cc-pwCV5Z
(e) Said difference is largest in molecules with significant degrees of static correlation, and negligible in molecules dominated by dynamical correlation
(f) The (mostly third-order) terms that make up the difference between CCSD-F12b and CCSD[F12] converge rapidly with the basis set, and are universally bonding. They can also be greatly reduced through basis set extrapolation.
(g) The fourth-order terms that make up the difference between CCSD(F12*) and CCSD[F12] converge more slowly, and are universally antibonding. These terms cannot be well reduced through basis set extrapolation, at least not in practical basis set regimes.
(h) The sometimes nonmonotonic basis set convergence of the CCSD(F12*)–CCSD-F12b difference results from the different rates of convergence of these last two opposing terms
(i) The CCSD-F12a approximation continues to have unacceptably large errors even near the basis set limit
(j) There is some evidence that, for n={5,6,7} the nZaPa family of basis sets exhibits somewhat smoother basis set convergence than the aug-cc-pV(n+d)Z sequence.
(k) If at all possible, CCSD(F12*) is to be preferred over CCSD-F12b. For codes that only have a closed-shell implementation of CCSD(F12*), such as MOLPRO, the use of closed-shell reaction cycles may represent a workaround.

## ACKNOWLEDGMENTS

The authors would like to thank Profs. Amir Karton (U. of Western Australia, Perth) and John F. Stanton (Quantum Theory Project, U. of Florida) for helpful discussions. JMLM would also like to thank Prof. Hans-Joachim Werner (U. Stuttgart, Germany) for helpful comments on a conference poster based on an early draft of this presentation.

This research was supported by the Israel Science Foundation (grant 1358/15) and by the Minerva Foundation, Munich, Germany, as well as by two internal Weizmann Institute funding sources: the Helen and Martin Kimmel Center for Molecular Design and a research grant from the estate of Emile Mimran. NS and MKK acknowledge doctoral and postdoctoral fellowships, respectively, from the Feinberg Graduate School (WIS).

## SUPPORTING INFORMATION

Our best reference MP2, MP2-F12, CCSD, and CCSD(F12*) data for W4-17 are freely available online (in Microsoft Excel format) in the FigShare data repository at http://doi.org/10.6084/m9.figshare.6818564, reference number 6818564.

# REFERENCES

(1) Ten-no, S. Explicitly Correlated Wave Functions: Summary and Perspective. *Theor. Chem. Acc.* **2012**, *131* (1), 1070 DOI: 10.1007/s00214-011-1070-1.

(2) Ten-no, S.; Noga, J. Explicitly Correlated Electronic Structure Theory from R12/F12 Ansätze. *Wiley Interdiscip. Rev. Comput. Mol. Sci.* **2012**, *2* (1), 114–125 DOI: 10.1002/wcms.68.

(3) Kong, L.; Bischoff, F. A.; Valeev, E. F. Explicitly Correlated R12/F12 Methods for Electronic Structure. *Chem. Rev.* **2012**, *112* (1), 75–107 DOI: 10.1021/cr200204r.

(4) Hättig, C.; Klopper, W.; Köhn, A.; Tew, D. P. Explicitly Correlated Electrons in Molecules. *Chem. Rev.* **2012**, *112* (1), 4–74 DOI: 10.1021/cr200168z.

(5) Vogiatzis, K. D.; Haunschild, R.; Klopper, W. Accurate Atomization Energies from Combining Coupled-Cluster Computations with Interference-Corrected Explicitly Correlated Second-Order Perturbation Theory. *Theor. Chem. Acc.* **2014**, *133* (3), 1–12 DOI: 10.1007/s00214-014-1446-0.

(6) Klopper, W.; Ruscic, B.; Tew, D. P.; Bischoff, F. A.; Wolfsegger, S. Atomization Energies from Coupled-Cluster Calculations Augmented with Explicitly-Correlated Perturbation Theory. *Chem. Phys.* **2009**, *356* (1–3), 14–24 DOI: 10.1016/j.chemphys.2008.11.013.

(7) Sylvetsky, N.; Peterson, K. A.; Karton, A.; Martin, J. M. L. Toward a W4-F12 Approach: Can Explicitly Correlated and Orbital-Based Ab Initio CCSD(T) Limits Be Reconciled? *J. Chem. Phys.* **2016**, *144* (21), 214101 DOI: 10.1063/1.4952410.

(8) Burns, L. A.; Marshall, M. S.; Sherrill, C. D.; Burns, L. A.; Marshall, M. S.; Sherrill, C. D. Appointing Silver and Bronze Standards for Noncovalent Interactions : A Comparison of Spin-Component-Scaled ( SCS ), Explicitly Correlated ( F12 ), and Specialized Wavefunction Approaches Appointing Silver and Bronze Standards for Noncovalent Interactions. *J. Chem. Phys.* **2014**, *141*, 234111 DOI: 10.1063/1.4903765.

(9) Sirianni, D. A.; Burns, L. A.; Sherrill, C. D. Comparison of Explicitly Correlated Methods for Computing High-Accuracy Benchmark Energies for Noncovalent Interactions. *J. Chem. Theory Comput.* **2017**, *13* (1), 86–99 DOI: 10.1021/acs.jctc.6b00797.

(10) Kesharwani, M. K.; Karton, A.; Martin, J. M. L. Benchmark Ab Initio Conformational Energies for the Proteinogenic Amino Acids through Explicitly Correlated Methods. Assessment of Density Functional Methods. *J. Chem. Theory Comput.* **2016**, *12* (1), 444–454 DOI: 10.1021/acs.jctc.5b01066.

(11) Brauer, B.; Kesharwani, M. K.; Kozuch, S.; Martin, J. M. L. The S66x8 Benchmark for Noncovalent Interactions Revisited: Explicitly Correlated Ab Initio Methods and Density Functional Theory. *Phys. Chem. Chem. Phys.* **2016**, *18* (31), 20905–20925 DOI: 10.1039/C6CP00688D.

(12) Manna, D.; Kesharwani, M. K.; Sylvetsky, N.; Martin, J. M. L. Conventional and Explicitly Correlated Ab Initio Benchmark Study on Water Clusters: Revision of the BEGDB and WATER27 Data Sets. *J. Chem. Theory Comput.* **2017**, *13* (7), 3136–3152 DOI: 10.1021/acs.jctc.6b01046.

(13) Tew, D. P.; Klopper, W.; Heckert, M.; Gauss, J. Basis Set Limit CCSD(T) Harmonic Vibrational Frequencies †. *J. Phys. Chem. A* **2007**, *111* (44), 11242–11248 DOI: 10.1021/jp070851u.

(14) Martin, J. M. L.; Kesharwani, M. K. Assessment of CCSD(T)-F12 Approximations and Basis Sets for Harmonic Vibrational Frequencies. *J. Chem. Theory Comput.* **2014**, *10* (5), 2085–2090 DOI: 10.1021/ct500174q.

(15) Tew, D. P.; Klopper, W.; Neiss, C.; Hättig, C. Quintuple-ζ Quality Coupled-Cluster Correlation Energies with Triple-ζ Basis Sets. *Phys. Chem. Chem. Phys.* **2007**, *9* (16), 1921–1930 DOI: 10.1039/B617230J.

(16) Ten-no, S. Initiation of Explicitly Correlated Slater-Type Geminal Theory. *Chem. Phys. Lett.* **2004**, *398* (1–3), 56–61 DOI: 10.1016/j.cplett.2004.09.041.

(17) Johnson, C. M.; Hirata, S.; Ten-no, S. Explicit Correlation Factors. *Chem. Phys. Lett.* **2017**, *683*, 247–252 DOI: 10.1016/j.cplett.2017.02.072.

(18) Shavitt, I.; Bartlett, R. J. *Many – Body Methods in Chemistry and Physics*; Cambridge University Press: Cambridge, 2009.

(19) Purvis, G. D.; Bartlett, R. J. A Full Coupled-cluster Singles and Doubles Model: The Inclusion of Disconnected Triples. *J. Chem. Phys.* **1982**, *76* (4), 1910–1918 DOI: 10.1063/1.443164.

(20) Raghavachari, K.; Trucks, G. W.; Pople, J. A.; Head-Gordon, M. A Fifth-Order Perturbation Comparison of Electron Correlation Theories. *Chem. Phys. Lett.* **1989**, *157* (6), 479–483 DOI: 10.1016/S0009-2614(89)87395-6.

(21) Watts, J. D.; Gauss, J.; Bartlett, R. J. Coupled-Cluster Methods with Noniterative Triple Excitations for

Restricted Open-Shell Hartree–Fock and Other General Single Determinant Reference Functions. Energies and Analytical Gradients. *J. Chem. Phys.* **1993**, *98* (11), 8718–8733 DOI: 10.1063/1.464480.
(22) Marchetti, O.; Werner, H.-J. Accurate Calculations of Intermolecular Interaction Energies Using Explicitly Correlated Coupled Cluster Wave Functions and a Dispersion-Weighted MP2 Method †. *J. Phys. Chem. A* **2009**, *113* (43), 11580–11585 DOI: 10.1021/jp9059467.
(23) Marchetti, O.; Werner, H.-J. Accurate Calculations of Intermolecular Interaction Energies Using Explicitly Correlated Wave Functions. *Phys. Chem. Chem. Phys.* **2008**, *10* (23), 3400–3409 DOI: 10.1039/b804334e.
(24) Peterson, K. A.; Kesharwani, M. K.; Martin, J. M. L. The Cc-PV5Z-F12 Basis Set: Reaching the Basis Set Limit in Explicitly Correlated Calculations. *Mol. Phys.* **2015**, *113* (13–14), 1551–1558 DOI: 10.1080/00268976.2014.985755.
(25) Köhn, A. Explicitly Correlated Connected Triple Excitations in Coupled-Cluster Theory. *J. Chem. Phys.* **2009**, *130* (13), 131101 DOI: 10.1063/1.3116792.
(26) Köhn, A. Explicitly Correlated Coupled-Cluster Theory Using Cusp Conditions. II. Treatment of Connected Triple Excitations. *J. Chem. Phys.* **2010**, *133* (17), 174118 DOI: 10.1063/1.3496373.
(27) Karton, A.; Daon, S.; Martin, J. M. L. W4-11: A High-Confidence Benchmark Dataset for Computational Thermochemistry Derived from First-Principles W4 Data. *Chem. Phys. Lett.* **2011**, *510*, 165–178 DOI: 10.1016/j.cplett.2011.05.007.
(28) Karton, A.; Sylvetsky, N.; Martin, J. M. L. W4-17: A Diverse and High-Confidence Dataset of Atomization Energies for Benchmarking High-Level Electronic Structure Methods. *J. Comput. Chem.* **2017**, *38* (24), 2063–2075 DOI: 10.1002/jcc.24854.
(29) Klopper, W.; Noga, J.; Koch, H.; Helgaker, T. Multiple Basis Sets in Calculations of Triples Corrections in Coupled-Cluster Theory. *Theor. Chem. Accounts Theory, Comput. Model. (Theoretica Chim. Acta)* **1997**, *97* (1–4), 164–176 DOI: 10.1007/s002140050250.
(30) Schwenke, D. W. The Extrapolation of One-Electron Basis Sets in Electronic Structure Calculations: How It Should Work and How It Can Be Made to Work. *J. Chem. Phys.* **2005**, *122* (1), 014107 DOI: 10.1063/1.1824880.
(31) Martin, J. M. L.; de Oliveira, G. Towards Standard Methods for Benchmark Quality Ab Initio Thermochemistry—W1 and W2 Theory. *J. Chem. Phys.* **1999**, *111* (5), 1843–1856 DOI: 10.1063/1.479454.
(32) Ranasinghe, D. S.; Petersson, G. A. CCSD(T)/CBS Atomic and Molecular Benchmarks for H through Ar. *J. Chem. Phys.* **2013**, *138* (14), 144104 DOI: 10.1063/1.4798707.
(33) Kalescky, R.; Kraka, E.; Cremer, D. Accurate Determination of the Binding Energy of the Formic Acid Dimer: The Importance of Geometry Relaxation. *J. Chem. Phys.* **2014**, *140* (8), 084315 DOI: 10.1063/1.4866696.
(34) Feller, D. Statistical Electronic Structure Calibration Study of the CCSD(T*)-F12b Method for Atomization Energies. *J. Phys. Chem. A* **2015**, *119* (28), 7375–7387 DOI: 10.1021/acs.jpca.5b00487.
(35) Miliordos, E.; Xantheas, S. S. On the Validity of the Basis Set Superposition Error and Complete Basis Set Limit Extrapolations for the Binding Energy of the Formic Acid Dimer. *J. Chem. Phys.* **2015**, *142*, 094311 DOI: 10.1063/1.4913766.
(36) Adler, T. B.; Knizia, G.; Werner, H.-J. A Simple and Efficient CCSD(T)-F12 Approximation. *J. Chem. Phys.* **2007**, *127* (22), 221106 DOI: 10.1063/1.2817618.
(37) Knizia, G.; Adler, T. B.; Werner, H.-J. Simplified CCSD(T)-F12 Methods: Theory and Benchmarks. *J. Chem. Phys.* **2009**, *130* (5), 054104 DOI: 10.1063/1.3054300.
(38) Hättig, C.; Tew, D. P.; Köhn, A. Communications: Accurate and Efficient Approximations to Explicitly Correlated Coupled-Cluster Singles and Doubles, CCSD-F12. *J. Chem. Phys.* **2010**, *132* (23), 231102 DOI: 10.1063/1.3442368.
(39) Hollett, J. W.; Gill, P. M. W. The Two Faces of Static Correlation. *J. Chem. Phys.* **2011**, *134* (11), 114111 DOI: 10.1063/1.3570574.
(40) Karton, A.; Tarnopolsky, A.; Lamère, J.-F.; Schatz, G. C.; Martin, J. M. L. Highly Accurate First-Principles Benchmark Data Sets for the Parametrization and Validation of Density Functional and Other Approximate Methods. Derivation of a Robust, Generally Applicable, Double-Hybrid Functional for Thermochemistry and Thermochemical . *J. Phys. Chem. A* **2008**, *112* (50), 12868–12886 DOI: 10.1021/jp801805p.
(41) Sylvetsky, N.; Martin, J. M. L. Probing the Basis Set Limit for Thermochemical Contributions of Inner-Shell Correlation: Balance of Core-Core and Core-Valence Contributions. *Mol. Phys.* **2018**, *EarlyView*, 1–10 DOI: 10.1080/00268976.2018.1478140.
(42) Peterson, K. A.; Dunning, T. H. Accurate Correlation Consistent Basis Sets for Molecular Core–Valence

(43) Hill, J. G.; Mazumder, S.; Peterson, K. A. Correlation Consistent Basis Sets for Molecular Core-Valence Effects with Explicitly Correlated Wave Functions: The Atoms B-Ne and Al-Ar. *J. Chem. Phys.* **2010**, *132* (5), 054108 DOI: doi:10.1063/1.3308483.

Correlation Effects: The Second Row Atoms Al–Ar, and the First Row Atoms B–Ne Revisited. *J. Chem. Phys.* **2002**, *117* (23), 10548–10560 DOI: 10.1063/1.1520138.

(44) Martin, J. M. L.; Parthiban, S. W1 and W2 Theories, and Their Variants: Thermochemistry in the KJ/Mol Accuracy Range. In *Quantum-Mechanical Prediction of Thermochemical Data*; Cioslowski, J., Ed.; Understanding Chemical Reactivity; Kluwer Academic Publishers: Dordrecht, 2002; Vol. 22, pp 31–65.

(45) Peterson, K. A.; Adler, T. B.; Werner, H.-J. Systematically Convergent Basis Sets for Explicitly Correlated Wavefunctions: The Atoms H, He, B-Ne, and Al-Ar. *J. Chem. Phys.* **2008**, *128* (8), 084102 DOI: 10.1063/1.2831537.

(46) Hill, J. G.; Peterson, K. A.; Knizia, G.; Werner, H.-J. Extrapolating MP2 and CCSD Explicitly Correlated Correlation Energies to the Complete Basis Set Limit with First and Second Row Correlation Consistent Basis Sets. *J. Chem. Phys.* **2009**, *131* (19), 194105 DOI: 10.1063/1.3265857.

(47) Feller, D.; Peterson, K. A.; Crawford, T. D. Sources of Error in Electronic Structure Calculations on Small Chemical Systems. *J. Chem. Phys.* **2006**, *124* (5), 054107 DOI: 10.1063/1.2137323.

(48) Dunning, T. H.; Peterson, K. A.; Wilson, A. K. Gaussian Basis Sets for Use in Correlated Molecular Calculations. X. The Atoms Aluminum through Argon Revisited. *J. Chem. Phys.* **2001**, *114* (21), 9244–9253 DOI: 10.1063/1.1367373.

(49) Martin, J. M. L. Heats of Formation of Perchloric Acid, HClO4, and Perchloric Anhydride, Cl2O7. Probing the Limits of W1 and W2 Theory. *J. Mol. Struct. THEOCHEM* **2006**, *771* (1–3), 19–26 DOI: 10.1016/j.theochem.2006.03.035.

(50) Zhong, S.; Barnes, E. C.; Petersson, G. A. Uniformly Convergent N-Tuple-ζ Augmented Polarized (NZaP) Basis Sets for Complete Basis Set Extrapolations. I. Self-Consistent Field Energies. *J. Chem. Phys.* **2008**, *129* (18), 184116 DOI: 10.1063/1.3009651.

(51) Barnes, E. C.; Petersson, G. A. MP2/CBS Atomic and Molecular Benchmarks for H through Ar. *J. Chem. Phys.* **2010**, *132* (11), 114111 DOI: 10.1063/1.3317476.

(52) Peterson, K. A. Chapter 11 Gaussian Basis Sets Exhibiting Systematic Convergence to the Complete Basis Set Limit. In *Annual Reports in Computational Chemistry*; 2007; Vol. 3, pp 195–206.

(53) Weigend, F.; Ahlrichs, R. Balanced Basis Sets of Split Valence, Triple Zeta Valence and Quadruple Zeta Valence Quality for H to Rn: Design and Assessment of Accuracy. *Phys. Chem. Chem. Phys.* **2005**, *7* (18), 3297–3305 DOI: 10.1039/b508541a.

(54) Frisch, M. J.; Trucks, G. W.; Schlegel, H. B.; Scuseria, G. E.; Robb, M. A.; Cheeseman, J. R.; Scalmani, G.; Barone, V.; Mennucci, B.; Petersson, G. A.; Nakatsuji, H.; Caricato, M.; Li, X.; Hratchian, H. P.; Izmaylov, A. F.; Bloino, J.; Zheng, G.; Sonnenberg, J. L.; Hada, M.; Ehara, M.; Toyota, K.; Fukuda, R.; Hasegawa, J.; Ishida, M.; Nakajima, T.; Honda, Y.; Kitao, O.; Nakai, H.; Vreven, T.; Montgomery, J. A., J.; Peralta, J. E.; Ogliaro, F.; Bearpark, M.; Heyd, J. J.; Brothers, E.; Kudin, K. N.; Staroverov, V. N.; Kobayashi, R.; Normand, J.; Raghavachari, K.; Rendell, A. P.; Burant, J. C.; Iyengar, S. S.; Tomasi, J.; Cossi, M.; Rega, N.; Millam, M. J.; Klene, M.; Knox, J. E.; Cross, J. B.; Bakken, V.; Adamo, C.; Jaramillo, J.; Gomperts, R.; Stratmann, R. E.; Yazyev, O.; Austin, A. J.; Cammi, R.; Pomelli, C.; Ochterski, J. W.; Martin, R. L.; Morokuma, K.; Zakrzewski, V. G.; Voth, G. A.; Salvador, P.; Dannenberg, J. J.; Dapprich, S.; Daniels, A. D.; Farkas, Ö.; Foresman, J. B.; Ortiz, J. V.; Cioslowski, J.; Fox, D. J. Gaussian 09 Rev. D01. Gaussian, Inc.: Wallingford, CT 2012.

(55) Frisch, M. J.; Trucks, G. W.; Schlegel, H. B.; Scuseria, G. E.; Robb, M. A.; Cheeseman, J. R.; Scalmani, G.; Barone, V.; Petersson, G. A.; Nakatsuji, H.; Li, X.; Caricato, M.; Marenich, A. V; Bloino, J.; Janesko, B. G.; Gomperts, R.; Mennucci, B.; Hratchian, H. P.; Ortiz, J. V; Izmaylov, A. F.; Sonnenberg, J. L.; Williams-Young, D.; Ding, F.; Lipparini, F.; Egidi, F.; Goings, J.; Peng, B.; Petrone, A.; Henderson, T.; Ranasinghe, D.; Zakrzewski, V. G.; Gao, J.; Rega, N.; Zheng, G.; Liang, W.; Hada, M.; Ehara, M.; Toyota, K.; Fukuda, R.; Hasegawa, J.; Ishida, M.; Nakajima, T.; Honda, Y.; Kitao, O.; Nakai, H.; Vreven, T.; Throssell, K.; Montgomery Jr., J. A.; Peralta, J. E.; Ogliaro, F.; Bearpark, M. J.; Heyd, J. J.; Brothers, E. N.; Kudin, K. N.; Staroverov, V. N.; Keith, T. A.; Kobayashi, R.; Normand, J.; Raghavachari, K.; Rendell, A. P.; Burant, J. C.; Iyengar, S. S.; Tomasi, J.; Cossi, M.; Millam, J. M.; Klene, M.; Adamo, C.; Cammi, R.; Ochterski, J. W.; Martin, R. L.; Morokuma, K.; Farkas, O.; Foresman, J. B.; Fox, D. J. Gaussian 16 Rev. B01. Gaussian, Inc.: Wallingford, CT 2016.

(56) Furche, F.; Ahlrichs, R.; Hättig, C.; Klopper, W.; Sierka, M.; Weigend, F. Turbomole. *Wiley Interdiscip.*

*Rev. Comput. Mol. Sci.* **2013**, *4* (2), 91–100 DOI: 10.1002/wcms.1162.
(57) Werner, H.-J.; Knowles, P. J.; Knizia, G.; Manby, F. R.; Schütz, M.; Celani, P.; Korona, T.; Lindh, R.; Mitrushenkov, A.; Rauhut, G.; Shamasundar, K. R.; Adler, T. B.; Amos, R. D.; Bernhardsson, A.; Berning, A.; Cooper, D. L.; Deegan, M. J. O.; Dobbyn, A. J.; Eckert, F.; Goll, E.; Hampel, C.; Hesselman, A.; Hetzer, G.; Hrenar, T.; Jansen, G.; Köppl, C.; Liu, Y.; Lloyd, A. W.; Mata, R. A.; May, A. J.; McNicholas, S. J.; Meyer, W.; Mura, M. E.; Nicklass, A.; O'Neill, D. P.; Palmieri, P.; Peng, D.; Pflüger, K.; Pitzer, R. M.; Reiher, M.; Shiozaki, T.; Stoll, H.; Stone, A. J.; Tarroni, R.; Thorsteinsson, T.; Wang, M. MOLPRO, Version 2015.1, a Package of Ab Initio Programs. University of Cardiff Chemistry Consultants (UC3): Cardiff, Wales, UK 2015.
(58) Karton, A.; Rabinovich, E.; Martin, J. M. L.; Ruscic, B. W4 Theory for Computational Thermochemistry: In Pursuit of Confident Sub-KJ/Mol Predictions. *J. Chem. Phys.* **2006**, *125* (14), 144108 DOI: 10.1063/1.2348881.
(59) Weigend, F.; Köhn, A.; Hättig, C. Efficient Use of the Correlation Consistent Basis Sets in Resolution of the Identity MP2 Calculations. *J. Chem. Phys.* **2002**, *116* (8), 3175–3183 DOI: 10.1063/1.1445115.
(60) Hättig, C. Optimization of Auxiliary Basis Sets for RI-MP2 and RI-CC2 Calculations: Core-Valence and Quintuple-Zeta Basis Sets for H to Ar and QZVPP Basis Sets for Li to Kr. *Phys. Chem. Chem. Phys.* **2005**, *7* (1), 59–66 DOI: 10.1039/b415208e.
(61) Weigend, F. A Fully Direct RI-HF Algorithm: Implementation, Optimised Auxiliary Basis Sets, Demonstration of Accuracy and Efficiency. *Phys. Chem. Chem. Phys.* **2002**, *4* (18), 4285–4291 DOI: 10.1039/b204199p.
(62) Yousaf, K. E.; Peterson, K. A. Optimized Complementary Auxiliary Basis Sets for Explicitly Correlated Methods: Aug-Cc-PVnZ Orbital Basis Sets. *Chem. Phys. Lett.* **2009**, *476* (4–6), 303–307 DOI: 10.1016/j.cplett.2009.06.003.
(63) Yousaf, K. E.; Peterson, K. A. Optimized Auxiliary Basis Sets for Explicitly Correlated Methods. *J. Chem. Phys.* **2008**, *129* (18), 184108 DOI: 10.1063/1.3009271.
(64) Martin, J. M. L. A Simple "Range Extender" for Basis Set Extrapolation Methods for MP2 and Coupled Cluster Correlation Energies. *AIP Conf. Proc.* **2018**, in press.
(65) Barnes, E. C.; Petersson, G. A.; Feller, D.; Peterson, K. A. The CCSD(T) Complete Basis Set Limit for Ne Revisited. *J. Chem. Phys.* **2008**, *129* (19), 194115 DOI: 10.1063/1.3013140.
(66) Varandas, A. J. C.; Pansini, F. N. N. Narrowing the Error in Electron Correlation Calculations by Basis Set Re-Hierarchization and Use of the Unified Singlet and Triplet Electron-Pair Extrapolation Scheme: Application to a Test Set of 106 Systems. *J. Chem. Phys.* **2014**, *141*, 224113 DOI: 10.1063/1.4903193.
(67) Pansini, F. N. N.; Neto, A. C.; Varandas, A. J. C. On the Performance of Various Hierarchized Bases in Extrapolating the Correlation Energy to the Complete Basis Set Limit. *Chem. Phys. Lett.* **2015**, *641*, 90–96 DOI: 10.1016/j.cplett.2015.10.064.
(68) Ranasinghe, D. S.; Frisch, M. J.; Petersson, G. A. A Density Functional for Core-Valence Correlation Energy. *J. Chem. Phys.* **2015**, *143* (21), 214111 DOI: 10.1063/1.4935973.
(69) Nyden, M. R.; Petersson, G. A. Complete Basis Set Correlation Energies. I. The Asymptotic Convergence of Pair Natural Orbital Expansions. *J. Chem. Phys.* **1981**, *75* (4), 1843 DOI: 10.1063/1.442208.
(70) Fliegl, H.; Klopper, W.; Hättig, C. Coupled-Cluster Theory with Simplified Linear-R12 Corrections: The CCSD(R12) Model. *J. Chem. Phys.* **2005**, *122* (8), 084107 DOI: 10.1063/1.1850094.
(71) Köhn, A.; Tew, D. P. Explicitly Correlated Coupled-Cluster Theory Using Cusp Conditions. I. Perturbation Analysis of Coupled-Cluster Singles and Doubles (CCSD-F12). *J. Chem. Phys.* **2010**, *133* (17), 174117 DOI: 10.1063/1.3496372.
(72) Tew, D. P. Explicitly Correlated Coupled-Cluster Theory with Brueckner Orbitals. *J. Chem. Phys.* **2016**, *145* (7) DOI: 10.1063/1.4960655.
(73) Valeev, E. F. Coupled-Cluster Methods with Perturbative Inclusion of Explicitly Correlated Terms: A Preliminary Investigation. *Phys. Chem. Chem. Phys.* **2008**, *10* (1), 106–113 DOI: 10.1039/b713938a.
(74) Hill, G.; Peterson, K. A.; Knizia, G.; Werner, H. Extrapolating MP2 and CCSD Explicitly Correlated Correlation Energies to the Complete Basis Set Limit with First and Second Row Correlation Consistent Basis Sets. *J. Chem. Phys.* **2009**, *131* (19), 194105.
(75) Noga, J.; Kutzelnigg, W. Coupled Cluster Theory That Takes Care of the Correlation Cusp by Inclusion of Linear Terms in the Interelectronic Coordinates. *J. Chem. Phys.* **1994**, *101* (9), 7738–7762 DOI: 10.1063/1.468266.
(76) Scuseria, G. E.; Tsuchimochi, T. Constrained-Pairing Mean-Field Theory. II. Exact Treatment of

Dissociations to Nondegenerate Orbitals. *J. Chem. Phys.* **2009**, *131* (16), 164119 DOI: 10.1063/1.3257965.
(77) Fogueri, U. R.; Kozuch, S.; Karton, A.; Martin, J. M. L. A Simple DFT-Based Diagnostic for Nondynamical Correlation. *Theor. Chem. Acc.* **2012**, *132* (1), 1291 DOI: 10.1007/s00214-012-1291-y.
(78) Ramos-Cordoba, E.; Salvador, P.; Matito, E. Separation of Dynamic and Nondynamic Correlation. *Phys. Chem. Chem. Phys.* **2016**, *18* (34), 24015–24023 DOI: 10.1039/C6CP03072F.
(79) Lee, T. J.; Taylor, P. R. A Diagnostic for Determining the Quality of Single-Reference Electron Correlation Methods. *Int. J. Quantum Chem.* **1989**, *36* (S23), 199–207 DOI: 10.1002/qua.560360824.
(80) Janssen, C. New Diagnostics for Coupled-Cluster and Møller-Plesset Perturbation Theory. *Chem. Phys. Lett.* **1998**, *290* (July), 423–430.
(81) Nielsen, I. M. B.; Janssen, C. L. Double-Substitution-Based Diagnostics for Coupled-Cluster and Møller–Plesset Perturbation Theory. *Chem. Phys. Lett.* **1999**, *310* (5–6), 568–576 DOI: 10.1016/S0009-2614(99)00770-8.